\documentclass{elsart}
\usepackage{cite}
\usepackage{subfigure}
\usepackage{epsfig}
\begin{document}
\def\bea{\begin{eqnarray}}   \def\eea{\end{eqnarray}}
\def \NN{$N\overline{N}~~$}
\def \DD{$D^0\overline{D^0}~~$}
\def \D0bar{$\overline{D^0}~~$}
\def \ee{$e^{+}e^{-}~~$}
\def \blankline{\par\vskip 12 pt\noindent}
\def\epem{ e^+e^- }
\def\c2{CLEO~II.V}
\def\ccb{ c\bar{c} }
\def\d0d0{ D^0\bar{D}^0 }
\def\p0p0{ P^0\bar{P}^0 }
\def\qp2{ \Bigl| \frac{q}{p} \Bigr|^2 }
\def\pq2{ \Bigl| \frac{p}{q} \Bigr|^2 }
\def\rarr{ \rightarrow }
\def\larr{ \leftarrow }
\def\jp{ J/\psi }
\def\pspr{ \psi^\prime }
\def\d0{D_2^{*0}}
\def\d+{D_2^{*+}}
\def\dst{D_2^*}
\def\mev{ \,{\rm MeV}/c^2  }
\def\gev{ \,{\rm GeV}/c^2  }
\begin{frontmatter}
 \begin{flushright}
  FERMILAB-Pub-03/411-E \\
  FRASCATI LNF-03/22(P)
 \end{flushright}
 \title{ 
 Measurement of Masses and Widths of
 Excited Charm Mesons $\dst$ and Evidence for Broad States  
 }
%
%
\author[ucd]{J.~M.~Link}
\author[ucd]{P.~M.~Yager}
\author[cbpf]{J.~C.~Anjos}
\author[cbpf]{I.~Bediaga}
\author[cbpf]{C.~G\"obel}
\author[cbpf]{A.~A.~Machado}
\author[cbpf]{J.~Magnin}
\author[cbpf]{A.~Massafferri}
\author[cbpf]{J.~M.~de~Miranda}
\author[cbpf]{I.~M.~Pepe}
% ...

\address[ucd]{University of California, Davis, CA 95616} 
\address[cbpf]{Centro Brasileiro de Pesquisas F\'\i sicas, Rio de Janeiro, RJ, Brasil} 
\address[cinv]{CINVESTAV, 07000 M\'exico City, DF, Mexico} 
\address[cu]{University of Colorado, Boulder, CO 80309} 
\address[fnal]{Fermi National Accelerator Laboratory, Batavia, IL 60510} 
\address[fras]{Laboratori Nazionali di Frascati dell'INFN, Frascati, Italy I-00044}
\address[guan]{University of Guanajuato, 37150 Leon, Guanajuato, Mexico} 
\address[ui]{University of Illinois, Urbana-Champaign, IL 61801} 
\address[iu]{Indiana University, Bloomington, IN 47405} 
\address[korea]{Korea University, Seoul, Korea 136-701}
\address[korea2]{Kyungpook National University, Taegu, Korea 702-701}
\address[milan]{INFN and University of Milano, Milano, Italy} 
\address[nc]{University of North Carolina, Asheville, NC 28804} 
\address[pavia]{Dipartimento di Fisica Nucleare e Teorica and INFN, Pavia, Italy} 
\address[pr]{University of Puerto Rico, Mayaguez, PR 00681} 
\address[sc]{University of South Carolina, Columbia, SC 29208} 
\address[ut]{University of Tennessee, Knoxville, TN 37996} 
\address[vu]{Vanderbilt University, Nashville, TN 37235} 
\address[wisc]{University of Wisconsin, Madison, WI 53706}

\endnote{\small See http://www-focus.fnal.gov/authors.html for
additional author information}

%
%
 \begin{abstract}
  
  Using data from the FOCUS experiment we analyze the 
  $D^+\pi^-$  and $D^0\pi^+$ 
  invariant mass distributions.
  We measure the $D_2^{*0}$
   mass 
  $M_{D_2^{*0}} = (2464.5 \pm 1.1 \pm 1.9) \mev$
  and width  
  $\Gamma_{D_2^{*0}} = (38.7 \pm 5.3 \pm 2.9) \mev$, 
  and the $D_2^{*+}$ mass 
  $ M_{D_2^{*+}} = (2467.6 \pm  1.5 \pm 0.76) \mev$ 
  and width
    $  \Gamma_{D_2^{*+}} =       (34.1  \pm 6.5 \pm 4.2) \mev$. 
  We find evidence for  broad structures
  over background in both the neutral and charged final state.
  If each is interpreted as evidence for a single $L=1$, $j_q=1/2$ excited charm
  meson resonance, the
  masses and widths are 
  $M_{1/2}^0 =(2407 \pm 21 \pm 35 ) \mev$,
  $\Gamma_{1/2}^0 = (240 \pm 55 \pm 59) \mev$,  
     and  $M_{1/2}^+ = (2403 \pm 14 \pm 35 ) \mev$ 
 $\Gamma_{1/2}^+ = (283 \pm 24 \pm 34 )  \mev$, respectively.
 \end{abstract}
\end{frontmatter}
  Interest in charm spectroscopy has shifted from the
 ground states of ($0^-$ and $1^-$) 
 $c\bar q$ mesons to the orbitally
 and radially excited states.
In the limit of infinitely heavy quark mass, 
 the heavy-light meson behaves analogously to the hydrogen atom, {\it i.e.,}
 the heavier quark does not contribute to the
 orbital degrees of freedom (which are completely defined by the light
 quark).
The angular momentum of the heavy quark is
 described by its spin $S_{Q}$, and  that of the light degrees of
freedom are described by  ${\bf j_{q}}\ = {\bf s_{q}}\, + \, {\bf L}$, 
where $s_q$ is the light quark spin and $L$ is the orbital
angular momentum of the light quark.
 The
quantum numbers $S_{Q}$ and $j_{q}$ are individually conserved. The quantum
numbers of the excited $L=1$ states are formed by combining 
${\bf S_Q}$ and ${\bf j_q}$. For $L=1$ we have $j_q=1/2$ and
    $j_q=3/2$. When combined with  $S_Q$ they provide  two
    $j_q=1/2$ (J=0,1) states, and two $j_q=3/2$ (J=1,2) states. 
    In this paper these
    four states will be denoted by 
    $D_0^*$, $D_1(j_q=1/2)$,
    $D_1(j_q=3/2)$ and $D_2^*$. Heavy Quark Symmetry (HQS) predicts the
    spectrum of excited charmed states 
    \cite{Shuryak:1981fz}-\cite{Eichten:1993ub}.
    In the HQS limit, conservation of both parity and $j_q$,
       requires that
       the strong decays $D_J^{(*)}(j_q=3/2) \to D^{(*)} \pi$ proceed only
       via a D-wave while the decays $D_J^{(*)}(j_q=1/2) \to D^{(*)} \pi$
       proceed only via an S-wave.  The states decaying to an
       S-wave are expected to be broad while those decaying in a D-wave are 
       known to be narrow \cite{reviews}\cite{pdg2002}.
       Models predict that, when the heavy quark is the charmed quark,
    the physical states will have properties very close to those
    of the heavy quark limit. 
  In the analysis described, we show the salient features of the $D^+\pi^-$ and $D^0 \pi^+$ 
  invariant mass distributions and measure parameters of the well-established 
 narrow states. We observe an excess of events in the mass 
 interval $2.25 \gev$ to $2.40 \gev$ that is consistent with a broad 
 resonance and must be included in the representation of the data to produce a good
 fit.
%
%
\par
The data for this paper were collected in the Wideband photoproduction
experiment FOCUS during the Fermilab 1996--1997 fixed-target
run. FOCUS
\cite{Link:2001pg}\cite{Link:2001dj}\cite{Link:2003ts} 
is an upgraded version of experiment
E687 \cite{Frabetti:1992au}\cite{Frabetti:1992bn}.
 In FOCUS, a forward multi-particle
spectrometer is used to  
investigate the interactions of high energy photons on a segmented BeO
target. We obtain a sample in excess of  1 million fully reconstructed
charm particles in three decay modes: $D^0 \rightarrow K^- \pi^+
,~K^- \pi^+ \pi^+ \pi^-$ and $D^+ \rightarrow K^- \pi^+ \pi^+$.  
(The charge-conjugate states are implicitly included throughout the paper.)
\par
The FOCUS detector is a large aperture, fixed-target spectrometer with
excellent vertexing and particle identification.
  A photon beam, with an endpoint energy of
 $\approx 300$ GeV,   is
derived from the bremsstrahlung of secondary electrons and positrons. The charged particles which emerge from
the target are tracked by two systems of silicon microvertex
detectors. The upstream system 
\cite{Link:2003ts}, consisting of 4 planes (two views in 2
stations), is interleaved with the experimental targets, while the
other system lies downstream of the target and consists of twelve
planes of microstrips arranged in three views. These detectors provide
high resolution separation of primary (production) and secondary
(decay) vertices with an average proper time resolution of $\approx
30~ {\rm fs}$ for 2-track vertices. The momentum of a charged particle
is determined with five stations of multiwire proportional
chambers by measuring deflections in two analysis magnets of
opposite polarity. Three multicell threshold \v Cerenkov counters
\cite{Link:2001pg}
 are used to discriminate between electrons, pions, kaons, and protons.  
\par
\section{Analysis Procedure and Results}
\begin{figure}[p]
  \begin{center}
   \subfigure
   {\includegraphics[scale=0.4]{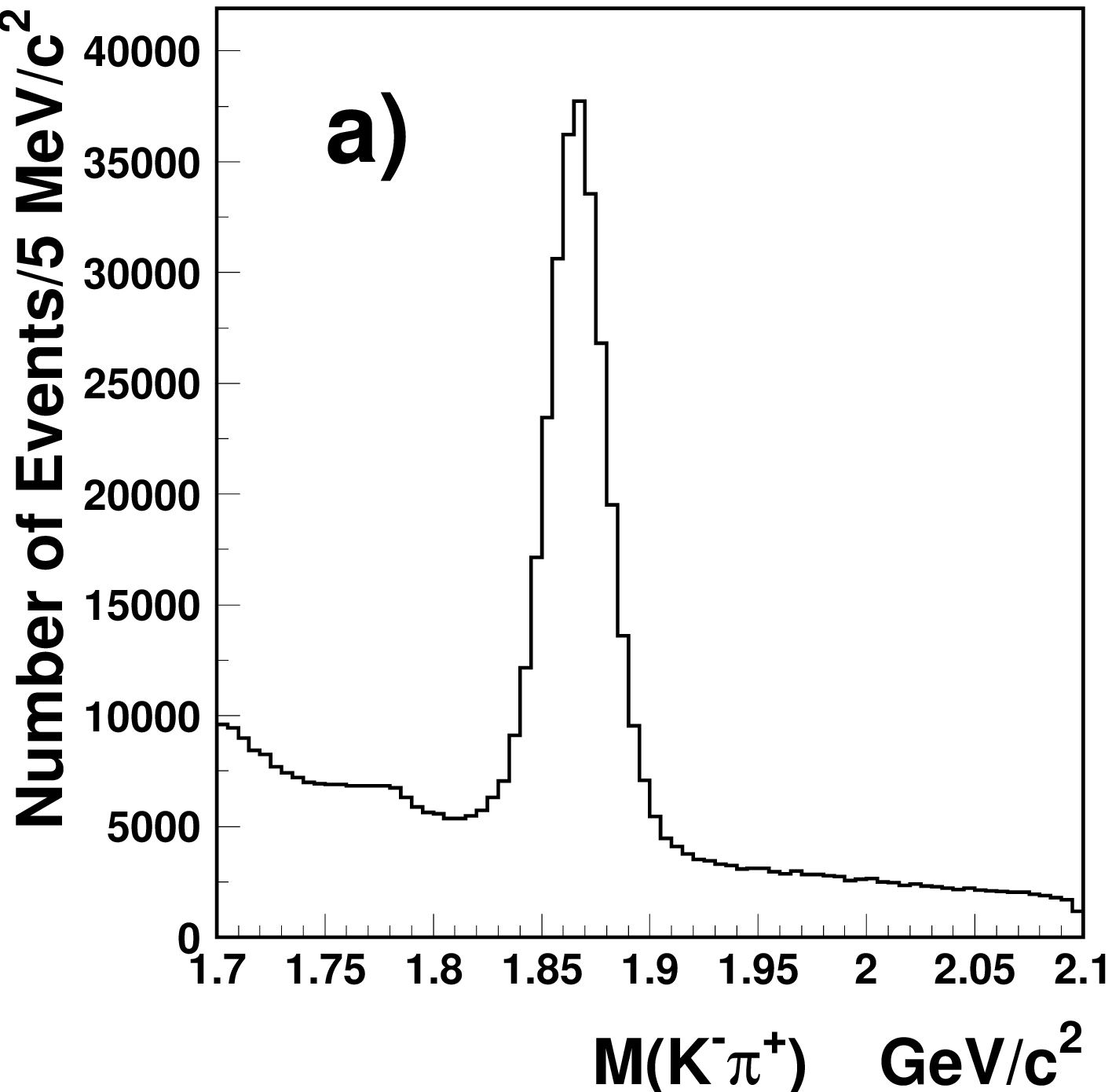}}
   \subfigure
   {\includegraphics[scale=0.4]{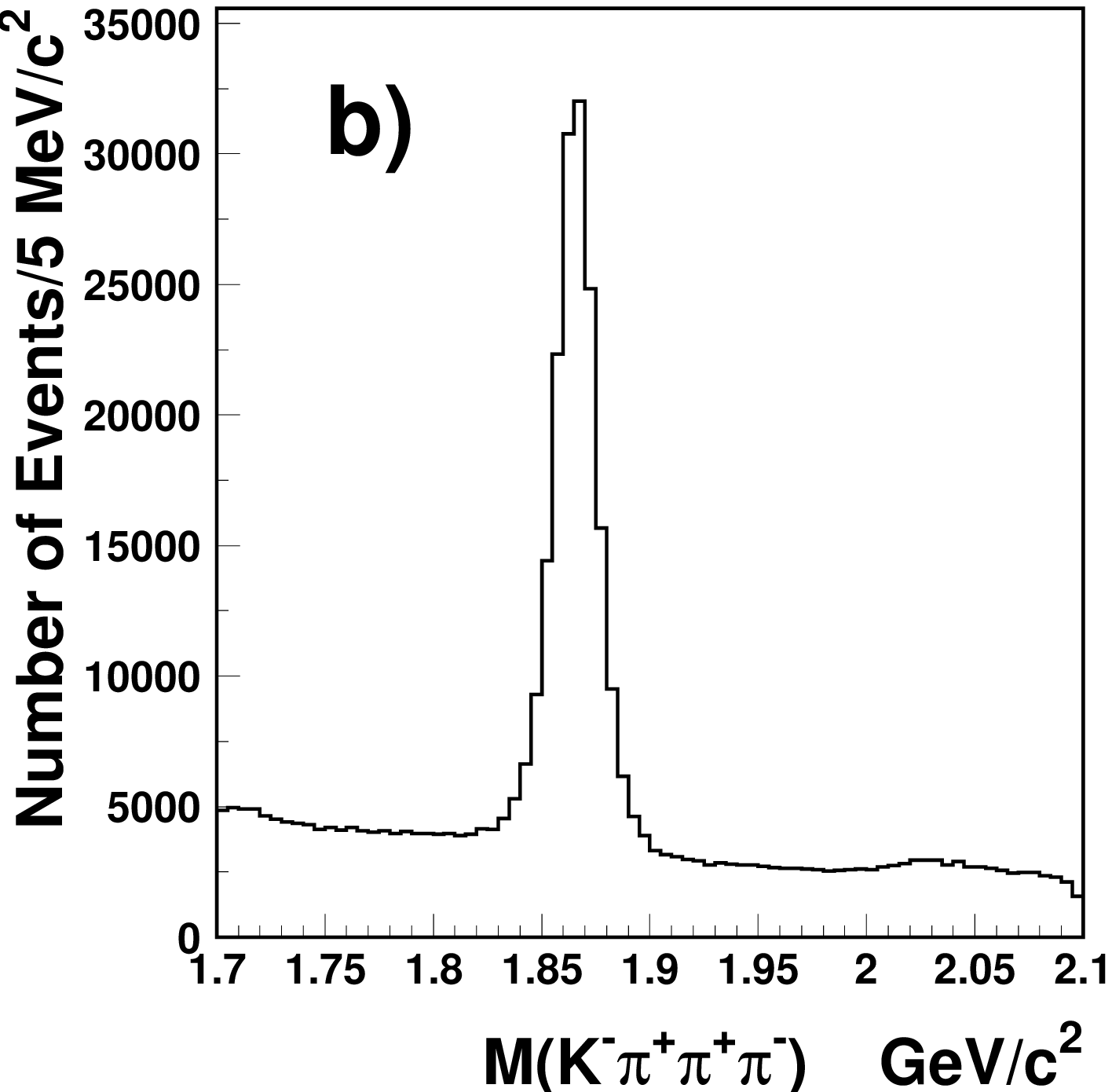}}
   \subfigure
   {\includegraphics[scale=0.4]{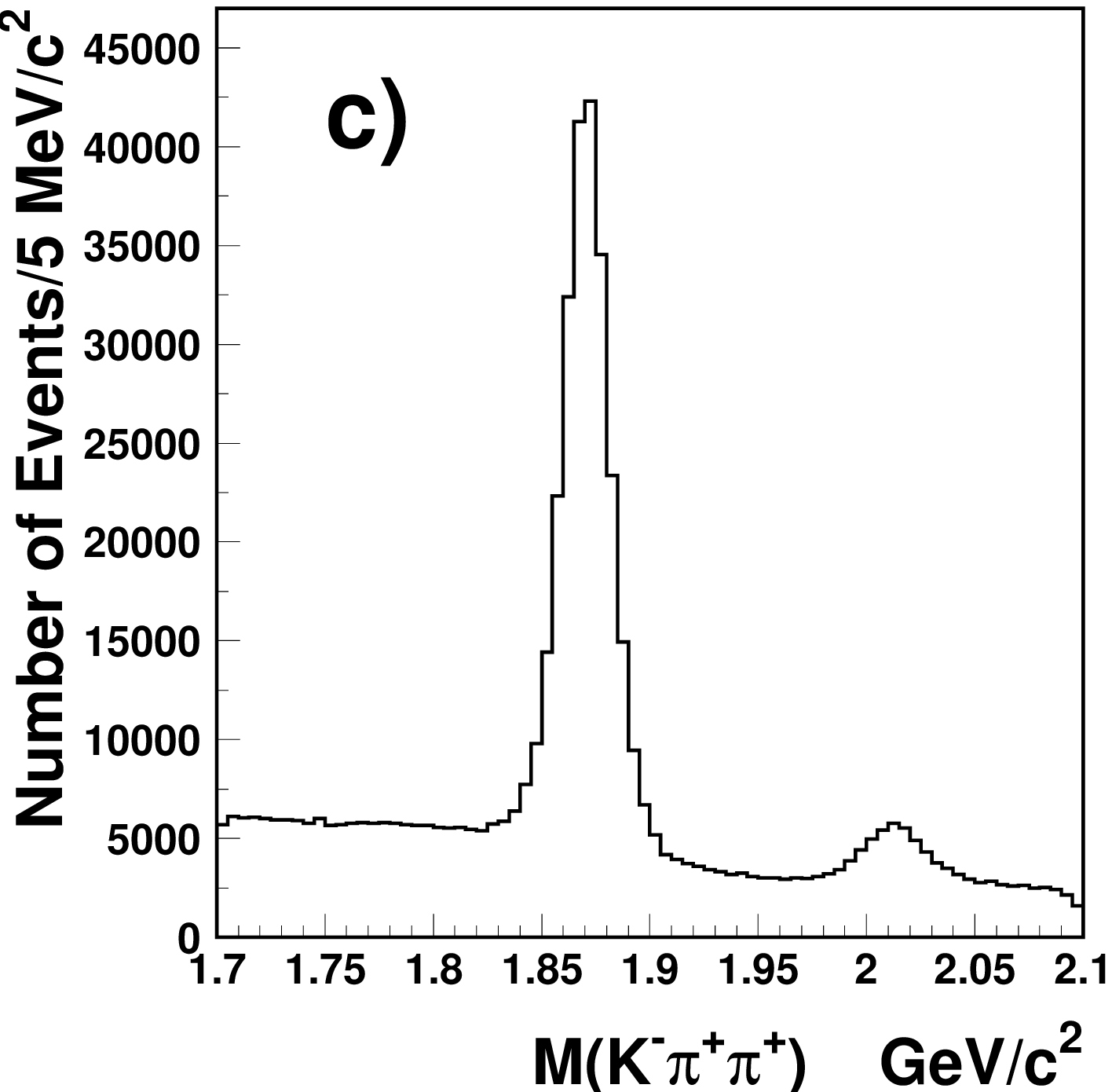}}
   \subfigure
   {\includegraphics[scale=0.4]{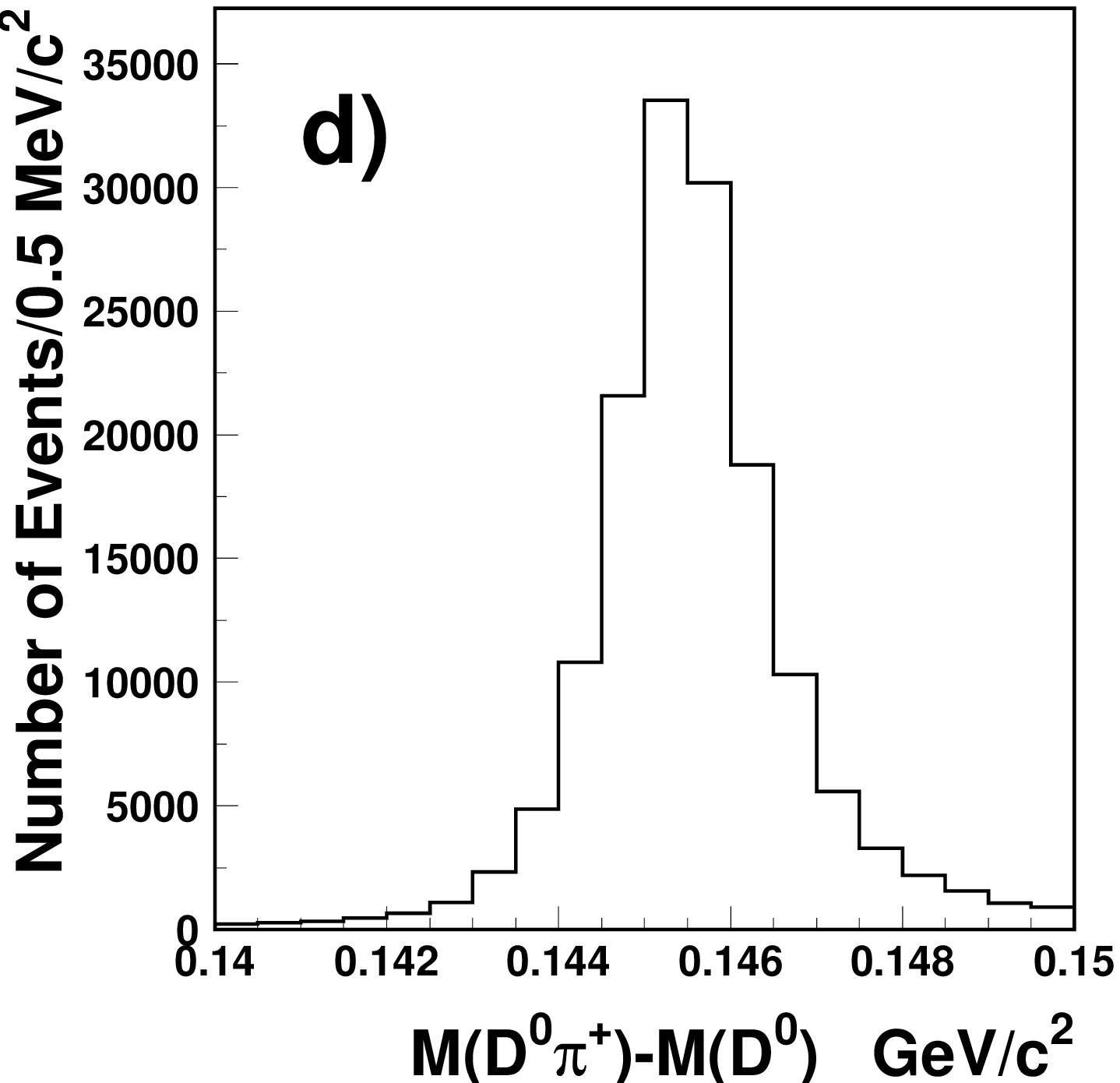}}
   \caption{Invariant mass plots for: 
        a)$D^0\rightarrow K^-\pi^+    $ ;
        b)$D^0\rightarrow K^-\pi^+ \pi^+ \pi^-$;
        c)$D^+\rightarrow K^-\pi^+ \pi^+$. 
     Invariant mass difference plot for d) $D^{*+} \rightarrow D^0\pi^+$. 
     }
   \label{fig:d0}
 \end{center}
\end{figure}

The $L=1$ charm mesons were reconstructed via
 $D^+\pi^-$ and $D^0\pi^+$ combinations.  The
    $D^0$ decays were reconstructed in the channels 
     $D^0\rightarrow K^- \pi^+ $ and
     $D^0\rightarrow K^- \pi^+ \pi^+ \pi^-$. The
$D^+$ decays were reconstructed in the channel 
  $D^+\rightarrow K^- \pi^+\pi^+$.
 To obtain a clean sample of high statistics charm decays,
    the vertexing and particle identification cuts were
    optimized separately for each decay mode. 
 The significance of separation between the
production and decay vertex, $\ell/\sigma_\ell$ (where $\ell$
 is the separation between the primary and
secondary vertex, and $\sigma_\ell$ is its error), 
was required to be
greater than 5, 10, and 12 respectively for the three decay modes.
 The interaction vertex was formed from the $D$ candidate, the
bachelor pion and at least one additional charged
 track \cite{Frabetti:1992au} and was
required to be located within the
target material. 
The pion and kaon candidates were required
to have a \v Cerenkov identification consistent with the selected
particle hypothesis. Further, we required that the decay 
$D^0\rightarrow K^- \pi^+ \pi^+ \pi^-$ 
be reconstructed
outside of target material and that $| \cos\theta_K| < 0.7$ for
the $D^0\rightarrow K^- \pi^+ $ decay, where $\theta_K$
is defined as the angle between the $D$ lab frame momentum and the 
kaon momentum in the $D$ center of mass frame. 
Our starting samples for the decay
modes with the above cuts are 210,000, 125,000 and 200,000 
events respectively (see Figure \ref{fig:d0}~a-c).  
Combinations within $\pm 2 \sigma$ of the nominal masses were retained
as $D$ candidates. Events with $D^0$  candidates
 coming from $D^{*+}$ decays were eliminated by applying 
 a $\pm 3 \sigma$ cut around the $D^{*+}-D^0$ mass difference 
 (see Figure \ref{fig:d0}~d).
 \par
\begin{figure}[p]
   \begin{center}
     {\includegraphics[scale=0.33]{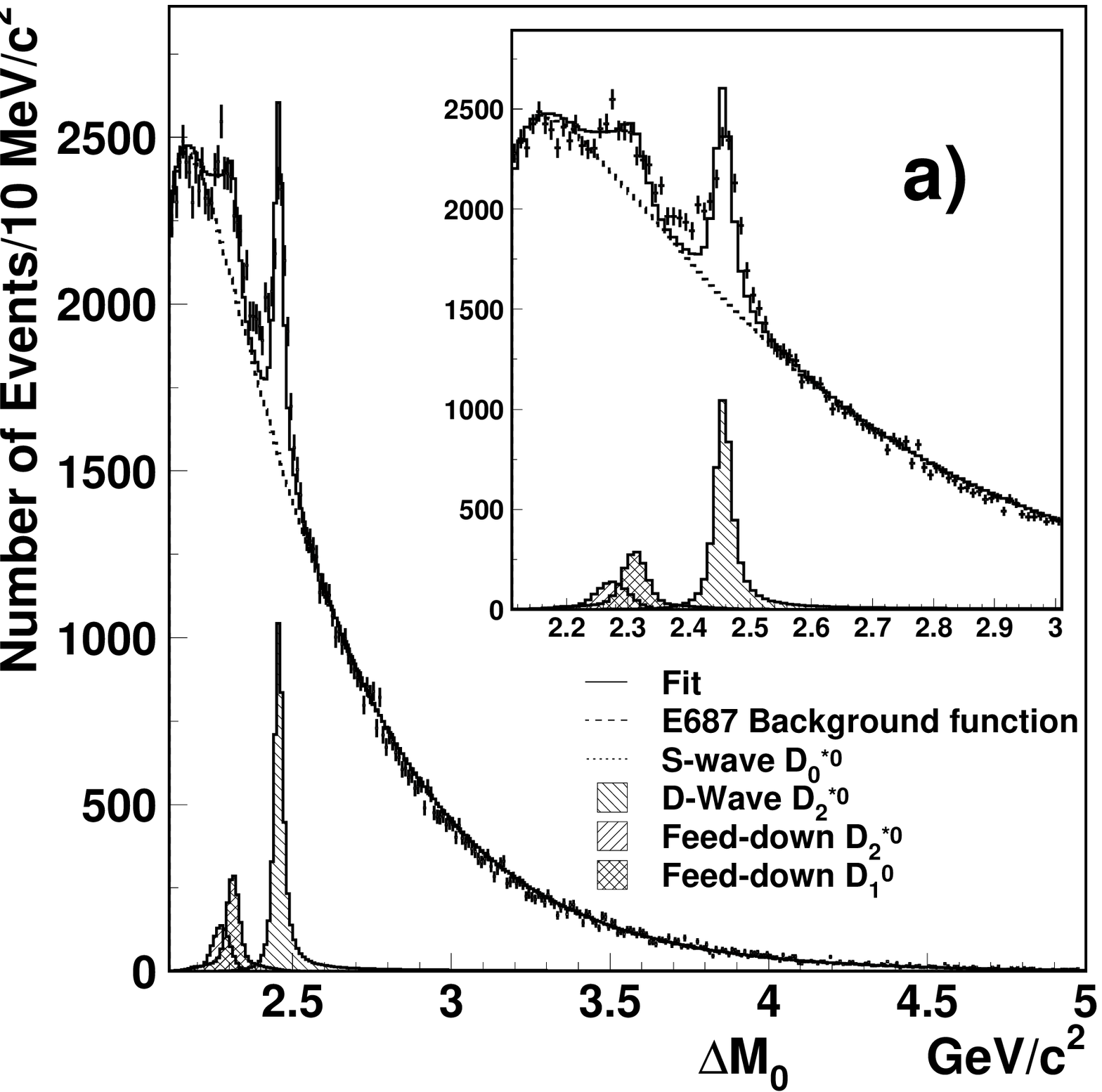}}
     {\includegraphics[scale=0.33]{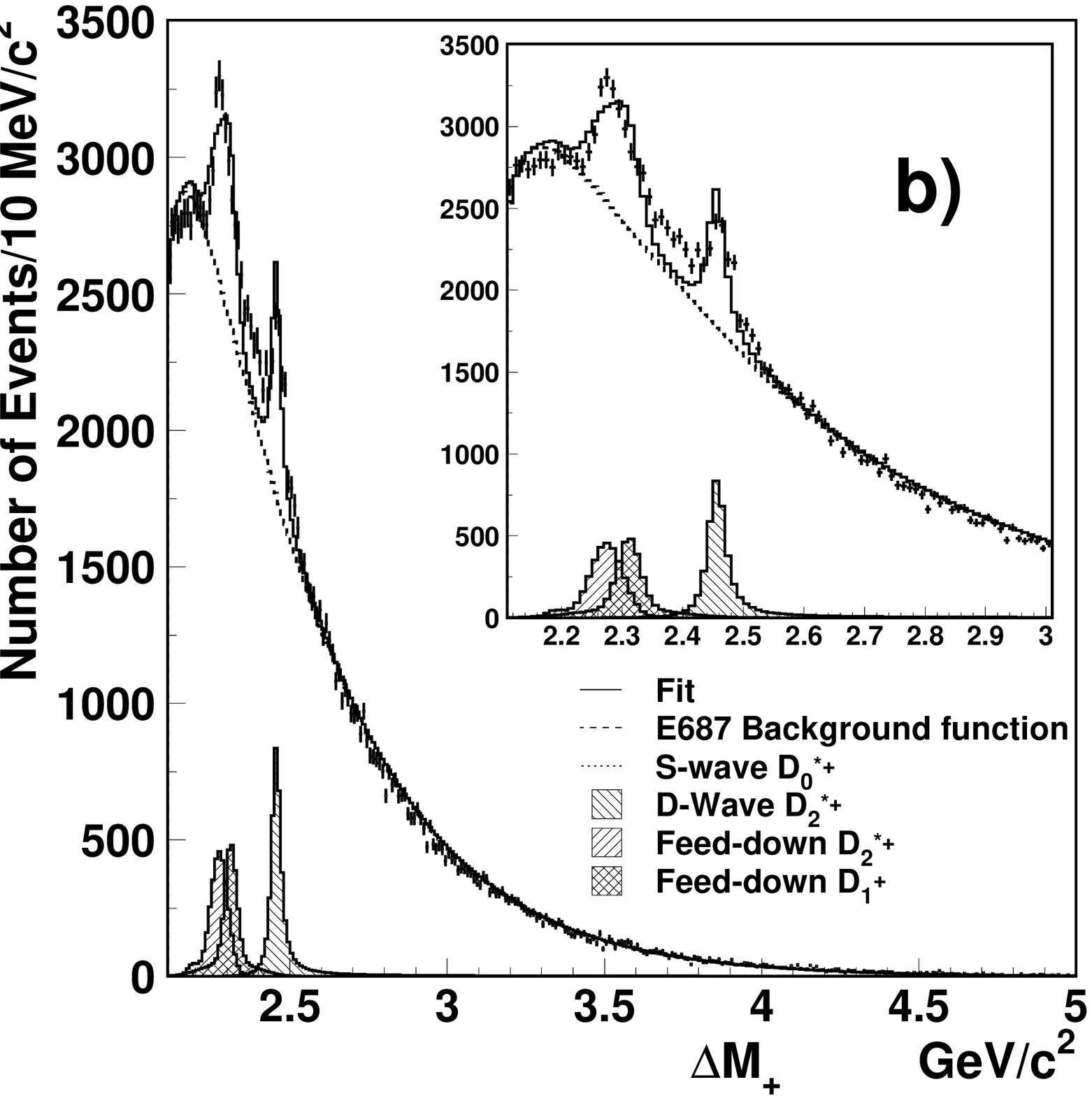}}
     {\includegraphics[scale=0.33]{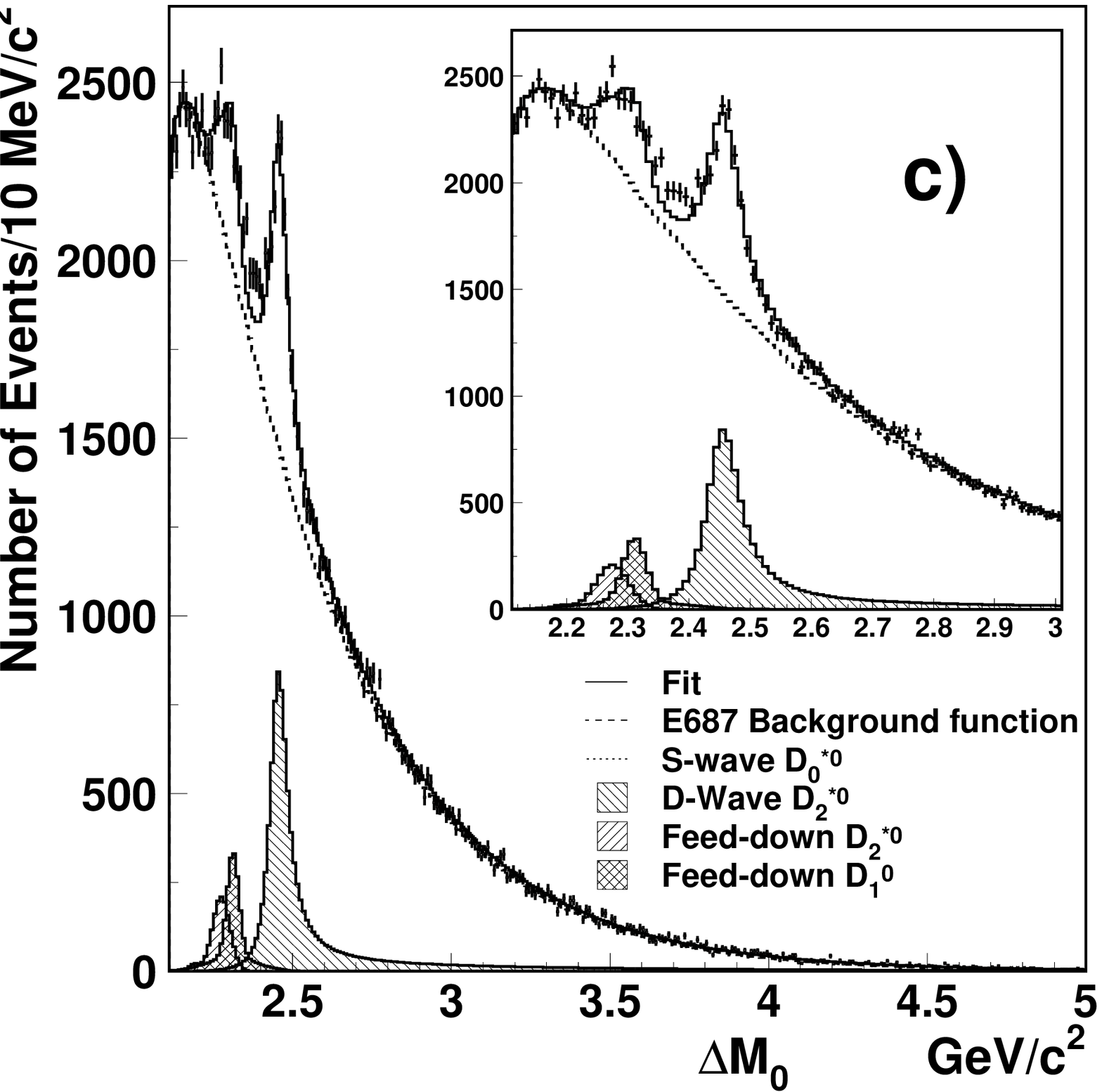}}
     {\includegraphics[scale=0.33]{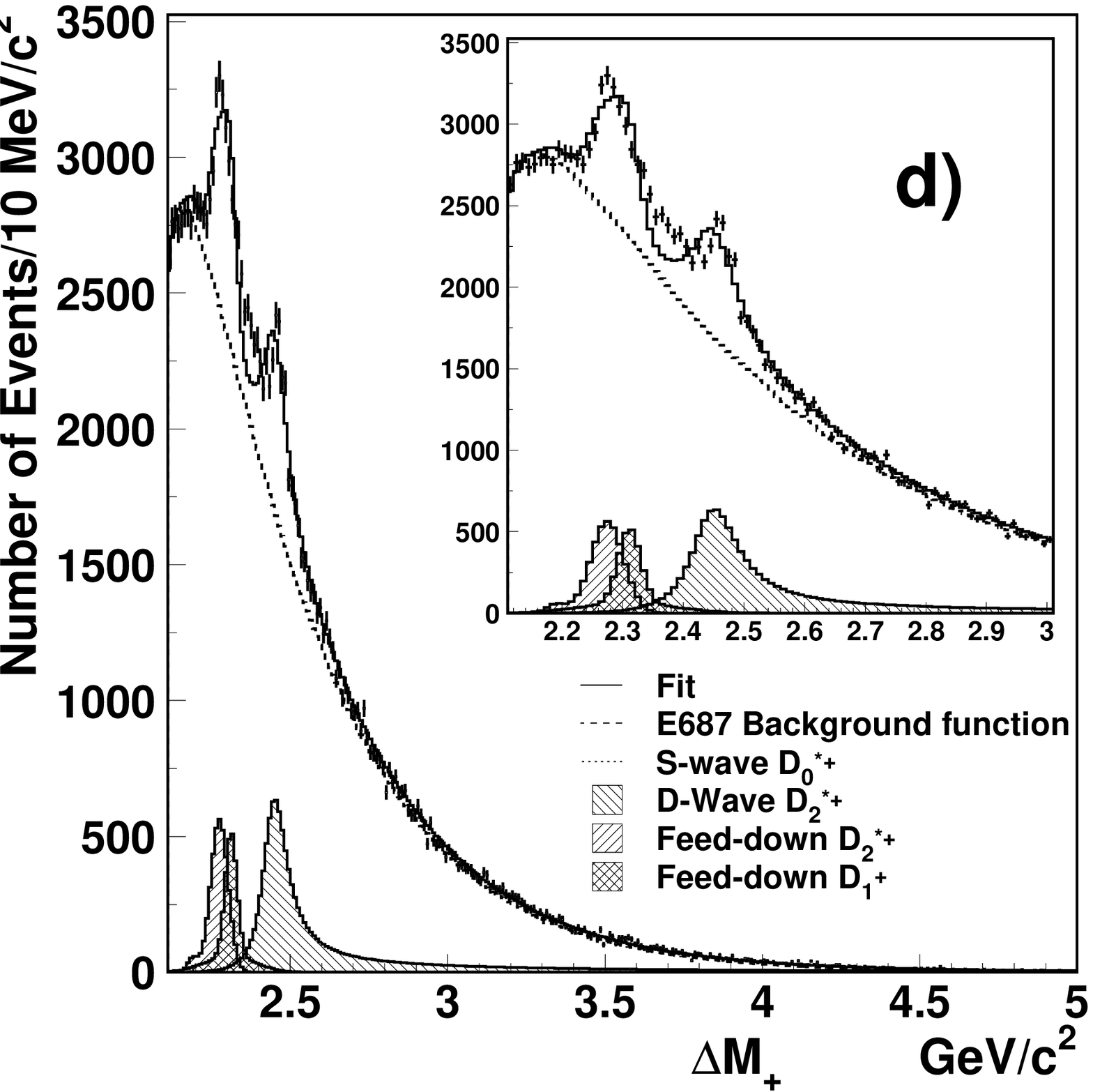}}
     \caption{
     The fit to the $D^+\pi^-$ and $D^0\pi^+$ mass spectra for the case where 
the D-wave mass and width are fixed to the PDG values, 
 the background is described by Equation \ref{eq:E687bkg},
and no broad resonance is included,
 is shown in a) and
b). The case where the D-wave mass and width are allowed to float in the
fit is shown in c) and d). 
Note that the none of these fits provides a good description of the data
between the feed-downs $(\sim 2300 \mev)$ and the $D_2^*$ peak $(\sim 
2500 \mev)$.
In Figure \ref{fig:d0+-fin}, 
we show that the data are well described when a broad resonance
is included in the fit.
       \label{fig:d0+-}}
   \end{center}
 \end{figure}

Figure~\ref{fig:d0+-}a) shows the distribution of the invariant mass
 difference 
 \begin{equation}
  \Delta M_0\equiv M((K^-\pi^+\pi^+)\pi^-) - M(K^-\pi^+\pi^+) + 
   M_{\mathrm{{PDG}}}(D^+)
 \end{equation}
 where $M_{\mathrm{{PDG}}}(D^+)$ is the world average $D^+$  mass  
  \cite{pdg2002}.
 Figure \ref{fig:d0+-}a) 
 shows a pronounced, narrow peak near a mass $M \approx 2460 \mev$,
   which is consistent with 
 the  $D_2^{*0}$ mass. 
  The
 additional enhancement at $M\approx 2300 \mev$ is consistent 
 with feed-downs from the states
 $D_1^0$ and $D_2^{*0}$ decaying to $D^{*+}\pi^-$ when 
 the $D^{*+}$ subsequently decays to a $D^+$ and undetected neutrals.
 \par
 The mass difference 
 \bea
   \Delta M_+ & \equiv &  M((K^-\pi^+,K^-\pi^+\pi^-\pi^+) \pi^+) - 
                M(K^-\pi^+,K^-\pi^+\pi^-\pi^+) +  \nonumber \\
              &        & + M_{\mathrm{PDG}}(D^0)
 \eea
 spectrum (Figure \ref{fig:d0+-}b) shows similar structures to the
 $\Delta M_0$ spectrum. The prominent peak is consistent with 
 a $D_2^{*+}$ of mass  $M \approx 2460 \mev$. 
 The additional enhancement at $M\approx 2300 \mev$ is again consistent
 with feed-downs. 
\par
 We fit the invariant mass difference histograms
 with terms for the $D_2^{*0},~D_2^{*+}$ peaks, 
 $D_1$ and $D^*_2$ feed-downs, combinatoric background and the possibility of a
 broad resonance. Fit terms were independent for each histogram 
 except for specific systematic tests, and all fit parameters were
 allowed to float except in tests which are described below. 
 
 The $D_2^{*0},D_2^{*+}$ signals were represented with relativistic D-wave
 Breit-Wigner functions 
  convoluted with a Gaussian resolution function
 $(\sigma = 7 \mev)$.
 The $\sigma$ of the resolution function was determined by
     processing PYTHIA \cite{Sjostrand:2003wg} 
     events through the FOCUS detector
     simulation and reconstruction codes.

The combinatoric background was represented by a continuum
      function discussed below.  The feed-downs were represented using
      line shapes determined by reconstructing simulated
      $D^*\pi$ events as $D \pi$.
 The masses
and widths used for the $D^*\pi$ and $D\pi$ came from the PDG or from our fit
to the $D_2^*$ as described below. Only the amplitudes of the feed-downs were
allowed to float in the fit.
 A 
 relativistic S-wave Breit-Wigner function was used to represent a
 broad resonance contribution (motivated below).
\par
In order to determine functions for the combinatoric background, several 
studies were performed. We studied the distribution of events
in wrong sign combinations (the $D^{*+}(D^0\pi^+)\pi^-$ reflection 
from the $D_1$ is very small), simulations where no $L=1$ charm mesons 
are present, and data sidebands of the $D^+$ and $D^0$. We found that
in all these cases, the combinatoric background is well described by a 
single exponential beyond $2250 \mev$. Several functions with 
threshold characteristics
(described in Section 2) were utilized to include information 
below $2250 \mev$.
Our final result is based on a function adapted from an E687
analysis \cite{Frabetti:1993vv} of excited $D$ states
 \begin{equation}
    \exp (A + B x) (x- C)^D
    \label{eq:E687bkg}
 \end{equation}  
where $x\equiv \Delta M_{0,+}$,
 and A, B, C and D are free parameters in the fit. (Care is required
to limit the range of the C parameter so that the threshold term does
not become imaginary.) With this function representing combinatoric
background, we produced final results that were stable with consistently good 
confidence levels over a variety of fit ranges. No combinatoric shapes
consistent with our background studies were able to describe either signal
histogram unless we included a function representing a broad resonance.
\par
In order to illustrate the motivation for including the broad resonance,
we show two representative fits performed without the broad resonance.
The distributions shown were fit with
 the $D_2^{*0},~D_2^{*+}$ parameters fixed to the world average values
 \cite{pdg2002} (Figure~\ref{fig:d0+-} a,b),
 and with the $D_2^{*0},~D_2^{*+}$
 parameters
 allowed to freely float (Figure~\ref{fig:d0+-} c,d).
  Individual fit components, and an expanded view of the
  region around $2400 \mev$ are shown in the figure.
  In both cases, the fit quality is unacceptable, 
  even 
  when the $D_2^*$ parameters float to values far from 
  the PDG values.
  Both fits indicate an excess of events between the $D_2^*$ signal and
  the feed-downs. We expect the background to be well described by a 
  single exponential in this region, but the fit is unable 
  to simultaneously describe the data at masses higher than 
  the $D_2^*$ peak and at masses lower than the $D_2^*$ peak.
  Since the behavior of the combinatoric background is heavily 
  influenced by the events with invariant mass difference 
  higher than  the $D_2^*$ peak, departures from the exponential form
  near $2400 \mev$ become evident.  
\par
While we are unable to rule out the possibility that
the excess is due to feed-down from higher mass charm states, 
we chose to describe the excess 
with an S-wave relativistic Breit-Wigner function
centered roughly near the excess.

\begin{figure}[p]
   \begin{center}
     {\includegraphics[scale=0.33]{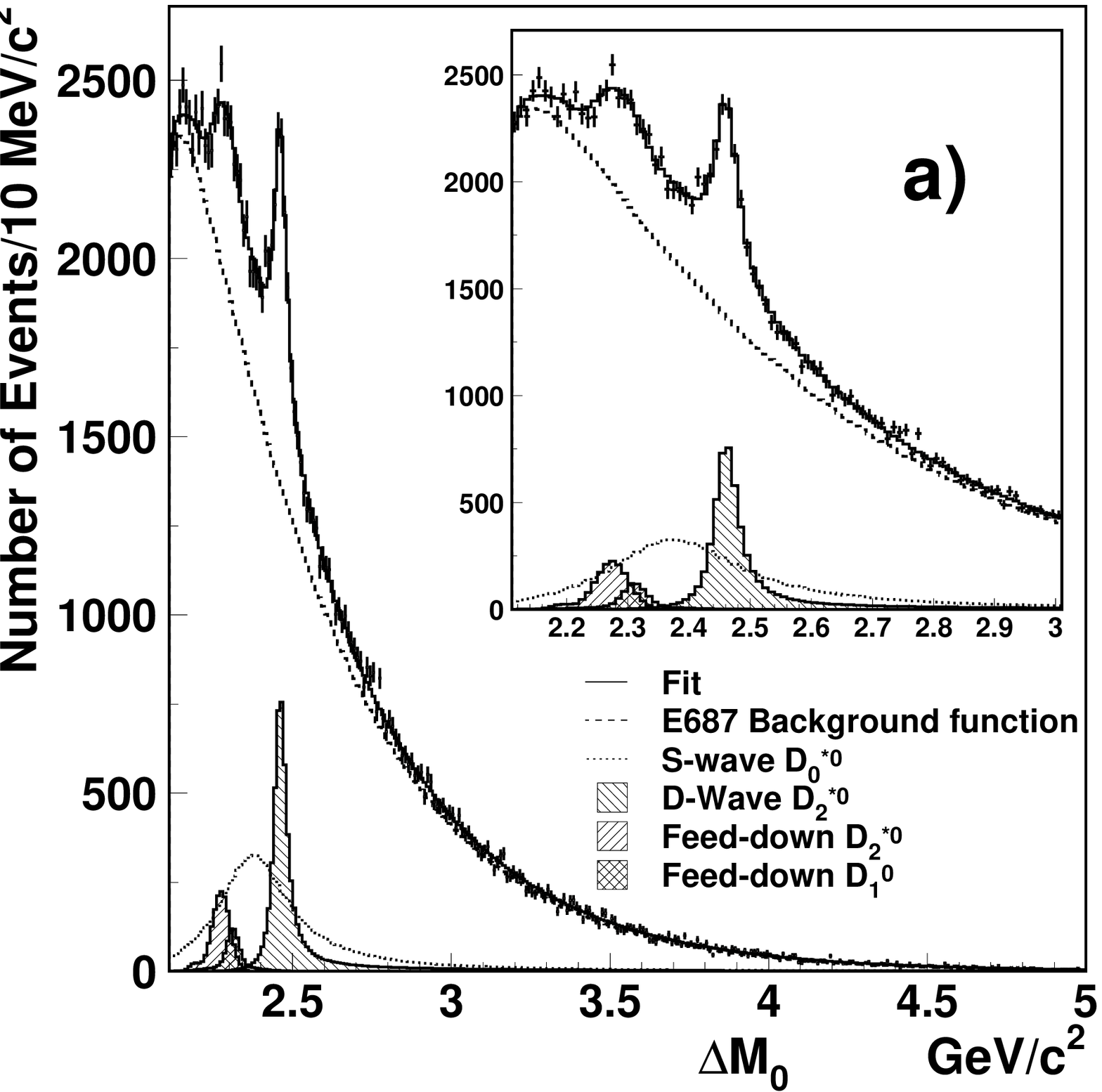}}
     {\includegraphics[scale=0.33]{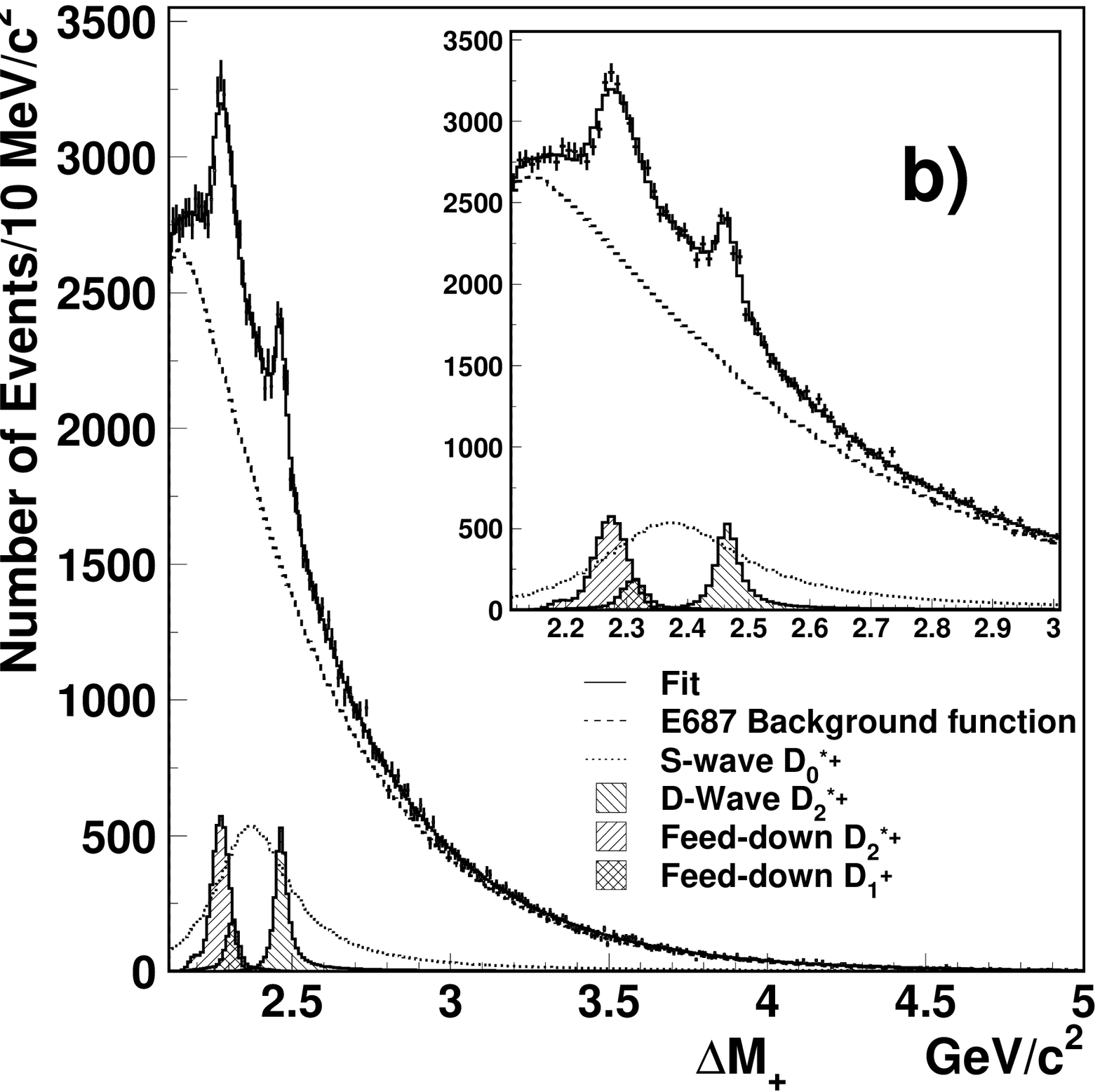}}  
     {\includegraphics[scale=0.33]{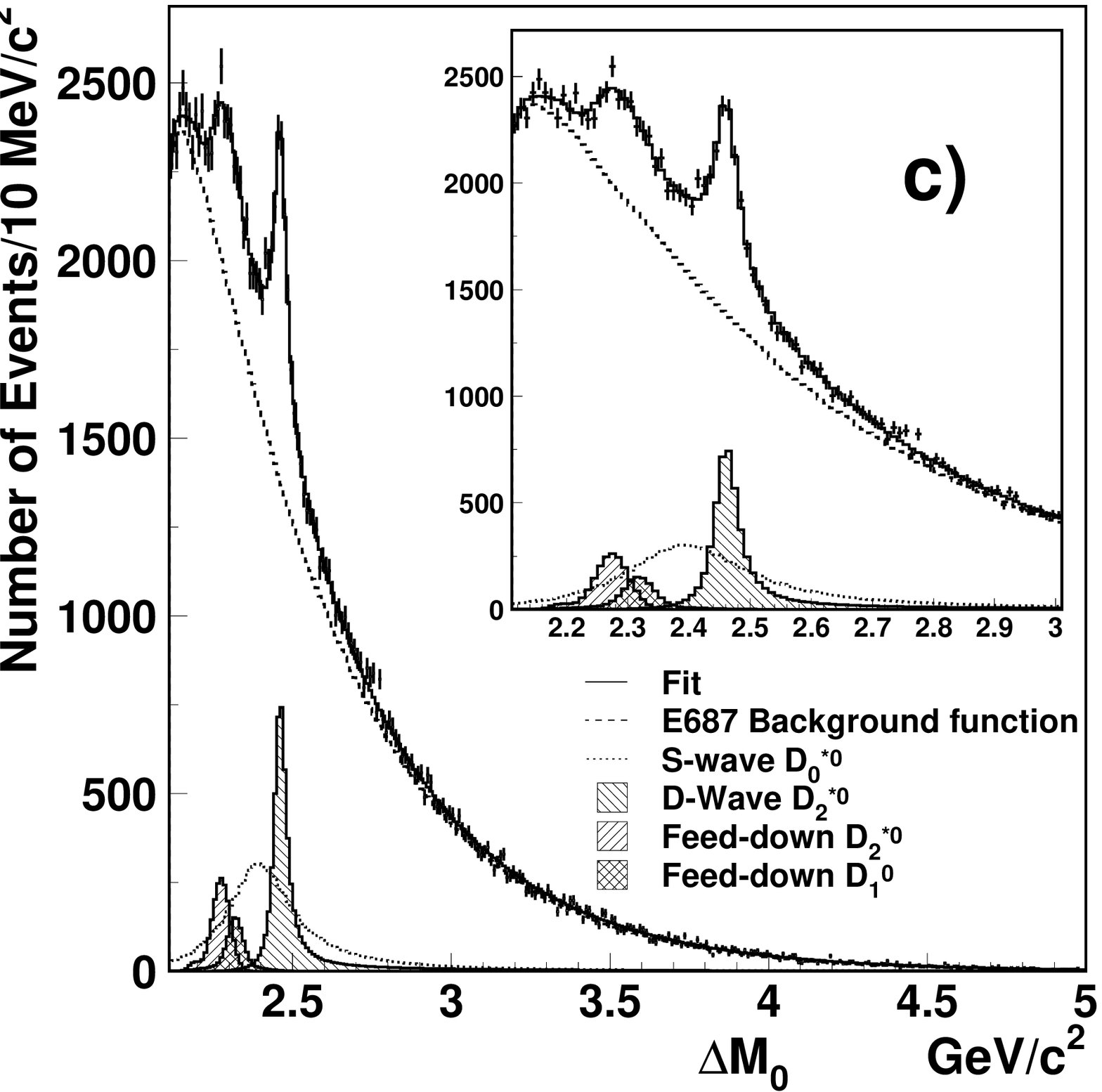}}
     {\includegraphics[scale=0.33]{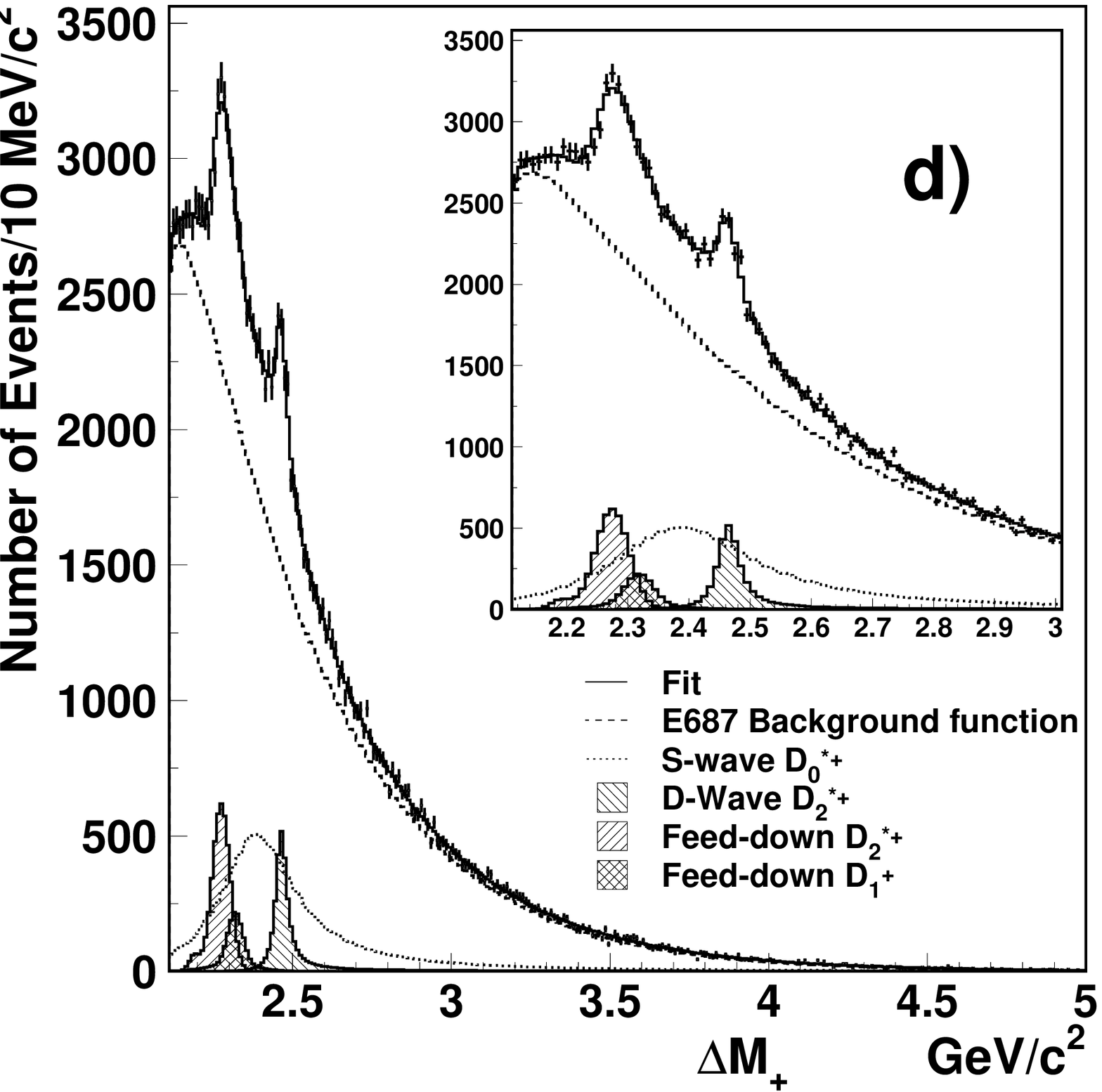}}          
     \caption{
     The fit to the $D^+\pi^-$ and $D^0\pi^+$ mass spectra including a term
 for an S-wave resonance. The case with the mass and width for the
 $D_1(3/2)$ and $D_2^*$ feed-downs fixed to the PDG values is shown in
 a) and b). The case with the mass and width for the $D_1(3/2)$ feed-down
 fixed to the PDG values and for the $D_2^*$
 feed-down determined by fits in a) and b) is shown in c) and d).  
 Notice the excellent agreement when the broad resonance is included
 (described in more detail in the text).
       \label{fig:d0+-fin}}
   \end{center}
 \end{figure}

In Figure \ref{fig:d0+-fin} a-b)  we show a fit to the data 
between $2120 \mev$  and
$5000 \mev$  that includes an S-wave relativistic Breit-Wigner in
addition to previously described terms. 
Agreement is excellent with a 
fit confidence level of $22$\%.
For self
consistency, the $D_2^*$ parameters measured in this fit are used to
recompute the feed-down lineshape. When the histograms were refit
using the new feed-down lineshape, the fit confidence level increased
to $28$\% without a significant change in the returned fit parameters.
     The results of this last fit are shown in Table \ref{tab:mass} 
     together with
PDG values where available.

We find that the mass and width returned by the fit are increased compared
to those reported by the PDG. Further, the yields and returned errors 
for the broad states indicate a significant  excess 
is present.

Although we are unable to distinguish between a broad state produced directly
via a $D^*_0$ and the feed-down from a broad $D_1$ state, we can make some
qualitative comparisons. If the ratio of $D_1$ to $D_2^*$ production is
the same for the charged and neutral modes, and the decays of these
states are dominated by $D\pi$ and/or $D^*\pi$, a meaningful comparison 
between the relative abundance of the feed-downs and the broad resonance to
the $D_2^*$ signal can be made.
With these assumptions, one expects that 
the feed-down from the
$D_2^*$ and $D_1$ narrow states to be larger 
relative to the $D_2^*$ peak for 
$D^0\pi^+$ modes since the $D^{*0}$ has no $D^{+}$ channel. 
This is  what we observe. We also find that the broad
state contribution in the $D^0\pi^+$ mode relative to the
$D_2^{*+}$ peak is larger than the broad
state contribution in the $D^+\pi^-$ relative to the 
$D_2^{*0}$ peak. This suggests some feed-down 
contribution to the broad state, perhaps from a broad
$D_1$ state (the search for a $D^*\pi$ broad resonance
is being performed and will be included in a later 
publication on $D^*\pi$ states).  

Further, the fit
parameters representing the broad S-wave state are statistically 
indistinguishable for both charged and neutral states. This is expected
for broad states differing only by the flavor of the light 
quark and dominated by decay into a $D^{(*)}$  meson and a pion.

\section{Systematic Checks}

Our systematic studies included a verification of the fit, fits 
using different functional forms
for the background, different shapes for the feed-down, fits excluding
the feed-down regions, fits over different regions of the data histogram,
a fit where we shifted our bin centers, a fit with the bin
size reduced by a factor of 2, fits in which we excluded  data where the background
shape is expected to differ from that of Equation \ref{eq:E687bkg}, and separate
fits for particle and anti-particle distributions. All the contributions were added in
quadrature (see Table \ref{tab:syserr}) and are described in more 
detail below.
\par
The fitting algorithm was extensively tested by fluctuating the 
data histogram, comparing errors returned by the fit and the spread 
of parameters from repeated trials.  We have also 
performed repeated fits to histograms generated with the fit function. 
We observe that the goodness of fit is acceptable, that the
     central values are unbiased and that the errors correctly
     describe the variation of the central values over the trials.
\par
We split the sample into particle and anti-particle, producing two 
statistically
separate data samples. These two samples were fit, and additional error 
(if any) was 
assessed until the parameters returned by the fit agreed with their 
average
\begin{equation}
  \sum (x-x_{\mathrm{avg}})^2 = \sigma_{\mathrm{stat}}^2+
   \sigma_{\mathrm{extra}}^2
\end{equation}
The deviations in the fit parameters returned by the tests described below
were added in quadrature to the split sample estimate to assess a total
systematic error.
\par
In addition to the modified E687 function, we
fit the data with a 
pure exponential background function
\begin{equation}
 \exp(A+Bx)
\end{equation}
We also fit the data with a
background function including a Gaussian term
\begin{equation}
  \exp(A+Bx+Cx^2)
\end{equation}
and we fit the data with a background function that was used by L3 \cite{L3bkg}
\begin{equation}
 \exp(A+Bx)/(1+\exp(D-x)/E)
\end{equation}
We used feed-down functions based on PDG values for the $D_2^*$ parameters.
In addition we used  feed-down functions based on 
our measured values for the $D_2^*$ parameters.
 We also fit the entire histogram from $2030 \mev$ 
to $5000 \mev$  while excluding the feed-down region ($2230 \mev  - 2400 \mev$)
with both the E687 modified function and the L3 function, and we performed 
 an 
additional
fit with the E687 modified function where we exclude the region between $2120 \mev$ 
and $2190 \mev$  in addition to excluding the feed-down regions.
\par
We find that the data samples at very high $(>30)$ $\ell/\sigma_\ell$ 
and 
high $D$ momentum $(P_D>70 {\rm GeV}/c)$
 have a significantly different background 
distribution. We test the effect on our final result by removing these
samples and by refitting.
\par
In order to determine the systematic uncertainty in our
mass difference due to the mass scale of the FOCUS spectrometer, 
we measured the mass differences $M(D^*)-M(D)$ and 
$M(\psi(2S)) - M(J/\psi)$. The quoted uncertainty is the additional 
contribution (added in quadrature to the statistical error)
needed for our measurements to be in agreement with 
world average values.
\par
\begin{figure}[p]
   \begin{center}
     \subfigure
     {\includegraphics[scale=0.33]{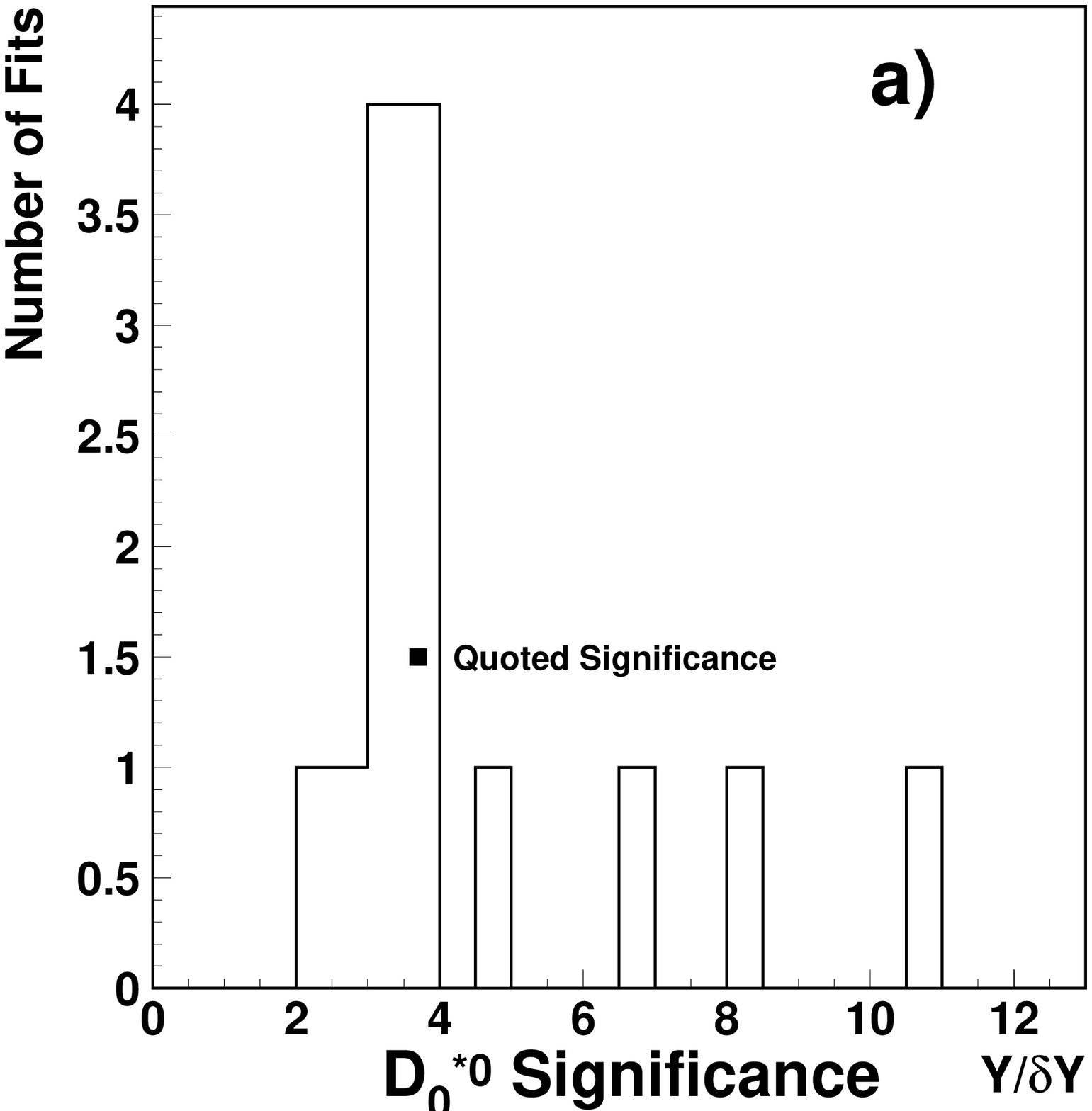}}
     \subfigure
     {\includegraphics[scale=0.33]{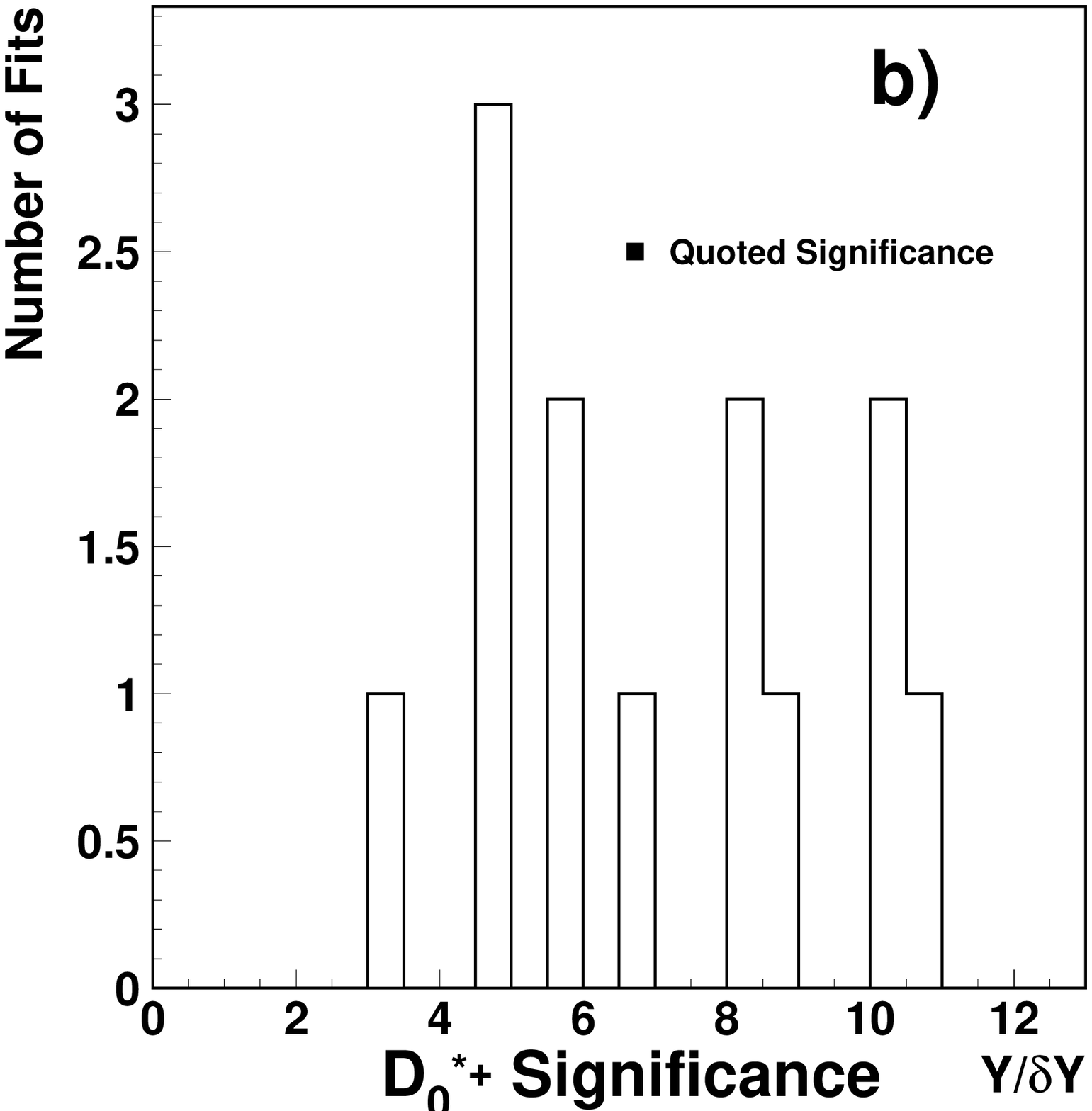}}
     \caption{Statistical significance of broad states 
      signals in the $D^+\pi^-$ (a) and $D^0\pi^+$ (b)
       channels for various fits.                                %
       \label{fig:nsigs}}
   \end{center}
 \end{figure}

 The contributions to the final systematic errors shown in Table \ref{tab:mass} 
 are listed in Table \ref{tab:syserr}.
 The yields for both the narrow and broad states 
  show a large variation depending on the fit considered. This
  is due to the wide range of background shapes investigated. 
 Further, since the broad resonance is not fully
contained in the fits, determination of the yield of the broad resonance
depends on how much of the data histogram is included in the fit, and
quoting a systematic error on this yield becomes problematic. Rather than
quote a systematic error on the yield of the broad state, we looked at the
statistical significance, $\mathrm{Yield}/\delta (\mathrm{Yield})$,
 for each fit considered. In
Figure  \ref{fig:nsigs}  we show that the statistical 
signficance of our quoted result is
a good representation of the fits tried.
  \par

 \begin{table}
     \caption{Measured masses and widths for narrow and broad 
     structures in $D^+\pi^-$ and
      $D^0\pi^+$ invariant mass spectra. The first error listed is statistical and the second
      is systematic. Units for the masses and widths are $\mev$. 
       \label{tab:mass}}

   \begin{center}
    \scriptsize
     \begin{tabular}{|l|c|c|c|c|c|} \hline
        & $D_2^{*0}$              & 
        $D_2^{*+}$             &
         $D_2^{*+} -  D_2^{*0}$ &
        $D_{1/2}^{0}$         &
         $D_{1/2}^{+}$ \\ 
                        &
                               &
                        &
                       &
                     &
                                 \\
       \hline
       Yield            & 
       $5776 \pm 869 \pm 696$  &
        $3474 \pm 670 \pm 656$      &
        &
         $  9810 \pm 2657$  &
         $18754 \pm 2189$\\
       \hline
       Mass             &
        $2464.5 \pm 1.1 \pm 1.9$ &
         $2467.6 \pm 1.5 \pm 0.76$   &
         $3.1 \pm 1.9 \pm 0.9$ &
          $2407 \pm  21 \pm 35$ &
           $2403\ \pm  14 \pm 35$\\
       PDG03          &
        $2458.9 \pm 2.0$         &
         $2459 \pm 4$                &
        $0 \pm 3.3$ &
          &
           \\
       \hline
       Width            &
        $38.7 \pm 5.3 \pm 2.9$   &
         $34.1 \pm 6.5 \pm 4.2$      &
                   &
          $240 \pm 55 \pm 59 $ &
           $283 \pm 24 \pm 34 $\\
       PDG03          &
        $23 \pm 5$               &
         $25^{+8}_{-7}$              &
             &
          &
          \\
       \hline
     \end{tabular}
    \vfill
   \end{center}
 \end{table}
 \begin{table}
   \caption{Individual contributions to the systematic error. Units are
    $\mev$}
   \label{tab:syserr}
   
   \scriptsize
   \begin{center}
     \begin{tabular}{|l|c|c|c|c|c|c|c|c|c|} \hline
       & $D_2^{*0}$ & $D_2^{*0}$  & $D_2^{*+}$ & $D_2^{*+}$      
       & $D_{1/2}^{0}$ & $D_{1/2}^{0}$  & $D_{1/2}^{+}$ & $D_{1/2}^{+}$ &
      $D_2^{*+} -  D_2^{*0}$ \\
       & Mass & Width & Mass & Width            
         & Mass & Width & Mass & Width                  &
         Mass   \\
       \hline
       $\ell/\sigma  < 30$        &
               0.160 &
                1.231 &
                        0.134 &
                                0.960                  &
                                        0.926 &
                                         15.73 &
                                          0.050 &
                                           2.871        &
                                            0.294          \\
       Part/Antipart                         &
               1.67 &
                 0 &
                          0.53 &
                                   0   &
                                        0 &
                                                 0 &
                                                          0 &
                  31.4  &
                          0 \\
       $P_D  < 70 {\rm GeV}/c $   & 
              0.227 &
                0.705 & 
                0.392 &
                        1.983                  &
                                2.482 &
                                 8.509 &
                                  10.38 &
                                   2.500  &
                                0.165              \\
       Different Fits                                &
               0.412 &
                0.272 &
                        0.124 &
                         0.693  &
                                10.48 &
                                 43.95 &
                                  1.439  &
                                        8.635  &
                                        0.353    \\
       Fit Regions                                & 
              0.376 &
                0.536 &
                        0.174 &
                                0.991  &
                                        1.571&
                                                12.80 & 
                                                1.209 &
                                                 6.657 &
                                                0.315              \\
       Feed-down tests       &
               0.633 &
                2.373 &
                        0.262 &
                                3.289                          &
                                        32.71 &
                                         31.91 &
                                          32.45 &
                                           6.137  &
                                           0.443           \\
       Binning  tests               &
                     0.442 &
                        0.576 &
                                0.113 &
                                        0.770                         &
                                                 6.584 &
                                                        6.652 &
                                                         6.380 &
                                                          0.894  &
                                                0.550         \\
       Mass Scale & 
       0.100 &
        0 &
         0.100 &
          0&
           0.100 &
            0& 
            0.100 & 
            0 &  
          0.100  \\
       \hline
       Total syst. error                & 
              1.94 &
                 2.89 & 
                   0.76 &
                           4.2          &
                                 35.1   &
                                 59.0 &
                                   34.7 &
                                     34.0 &
                0.91                               \\
       \hline
     \end{tabular}
     \vfill
   \end{center}
 \end{table}

 \par
 \begin{table}
   \caption{Predicted mass differences with respect to the $D$ meson
     compared to this result.
     The charged and neutral states are averaged.
     In the case of the broad state
      we compare our result to $D_0^*$ only. Units are $\mev$. 
     \label{tab:thmass}
}
   \begin{scriptsize}
     \begin{center}
       \begin{tabular}{|l|c|c|c|c|} \hline
         Reference      &
           $D_2^*$ &
            $D_1 $ &
             $D_1 $  &
             $D_0^*$                   \\
                         ~     &
         $j_q=3/2$   &
            $j_q=3/2$ &
             $j_q=1/2$  &
                         $j_q=1/2$     \\
         ~         &
          $~^3P_2$ &
           $~^3P_1$        &
            $~^1P_1$         &
            $~^3P_0$                   \\ 
         \hline\hline
         This paper      &
          599    &
                 &
                       &
                   538   \\
         World Av. \cite{pdg2002} &
          593   &
           556   &
                 &
                         \\
   Kalashnikova et al. (2002) \cite{Kalashnikova:2001px}
        & 579    & 562   & 603  &  564   \\
           Di~Pierro et al. (2001) \cite{DiPierro:2001uu}
     & 592    & 549   &  622    &  509      \\
   Ebert  et al. (1998) \cite{Ebert:2001zm}
       & 584    & 539   & 626  &  563   \\
   Isgur  (1998) \cite{Isgur:kr}
        & 594    & 549   & 719  &  699   \\
   Godfrey and Kokoski (1991)  \cite{Godfrey:wj}
 & 620    & 590   & 580  &  520   \\
   Godfrey and Isgur (1985) \cite{Godfrey:xj}
      & 620    & 610   & 560  &  520   \\
   Eichten et al. (1980) \cite{Eichten:1979ms}
      & 645    & 637   & 498  &  489   \\
   Barbieri et al. (1976) \cite{Barbieri:1975jd}
      & 428    & 380   & 339  &  259   \\
   De~Rujula et al. (1976) \cite{DeRujula:1976kk}
      & 494    & 464   & 384  &  374   \\   
   \hline\hline
       \end{tabular}
       \vfill
     \end{center}
   \end{scriptsize}
 \end{table}
 \section{Conclusions}
 FOCUS has measured the $D^+\pi^-$ and  $D^0\pi^+$ mass
 spectra and provided new values for the masses and widths of the
 $D_2^{*0}$ and $D_2^{*+}$  mesons (Table \ref{tab:mass}) 
 with errors less than  or equal to
 the errors on world averages.
\par
 The $\dst$ masses and widths measured are found to be higher than 
 the world averages. We
 attribute the change  
 to the inclusion   of an underlying broad state.
 
 We find significant evidence for a broad excess which
 we parameterize with an S-wave resonance. Our results are consistent
 with a broad resonance occuring near $2400 \mev$ with a width
 of about $250 \mev$ in both the charged and neutral modes. We are
 unable to distinguish whether the broad excess is due to a state
 such as the $D^*_0$, predicted by HQS at $M\approx 2400 $ 
 and width $\approx 100-200 \mev$,
 or due to feed-down from another broad state, such as the $D_1(j_q=1/2)$, or
 whether both states contribute.
 
 Evidence for $L=1$ broad (S-wave) states has been previously presented 
 in B decays by CLEO in the $D^{*+}\pi^-$ final state \cite{Anderson99},
 and BELLE \cite{Abe:2003zm} in the $D^{*+}\pi^-, D^+\pi^-$ final states.
\par
 Our measurements are compared to theory predictions in
 Table \ref{tab:thmass}. The $\dst$  masses are in good agreement with
  \cite{Isgur:kr} and \cite{DiPierro:2001uu}.
  Reference \cite{DiPierro:2001uu}, in addition,  predicts a 
 $\dst$-$D_0^*$ mass shift consistent with our evidence, while  
 \cite{Isgur:kr} predicts a shift with the opposite sign.
\par
\vspace{1.0cm}
 We would like to thank the staffs of Fermi National
 Accelerator Laboratory, INFN of Italy, and the physics departments
 of the collaborating institutions for their assistance. 
 This research was partly supported 
 by the U.~S. Department of
 Energy, the U.~S. National Science Foundation, 
 the Italian Istituto
 Nazionale di Fisica Nucleare and 
 Ministero dell'Istruzione, dell'Universit\`a e della Ricerca,
 the Brazilian Conselho Nacional de Desenvolvimento Cient\'{\i}fico e
 Tecnol\'ogico, CONACyT-M\'exico, the Korean Ministry of Education, and
 the Korean Science and Engineering Foundation. 
\end{document}